\documentclass[twocolumn,showpacs,preprintnumbers,footinbib%
               aps,prd]{revtex4-1}
\usepackage[dvips]{graphicx}
\usepackage{amsmath,amssymb}
\usepackage[breaklinks, colorlinks=true, pdfstartview=FitV, linkcolor=red,%
 citecolor=blue, urlcolor=blue]{hyperref}

\newcommand{\Nc}{N_{\text{c}}}
\newcommand{\Nf}{N_{\text{f}}}
\newcommand{\muq}{\mu_{\text{q}}}

\begin{document}

\title{Generic features of the phase transition in cold and dense
  quark matter}

\author{Kenji Fukushima}
\affiliation{Department of Physics, Keio University,
             Kanagawa 223-8522, Japan}

\begin{abstract}
  We investigate the phase transition in cold and dense quark matter
  in an intuitive way that shares common features of the effective
  model approaches.  We first express the quasi-particle contribution
  to the thermodynamic potential with the dynamical mass $M$ and then
  discuss how we can understand the possible first-order phase
  transition with and without the vector interaction from the
  saturation curve on the plane of the energy per particle and the
  density.  We next extend our analysis including inhomogeneity and
  discuss the relation between the order of the phase transition and
  the saturation curve.
  We emphasize that the saturation curve is useful to infer
  qualitative nature of the phase transition even without knowing the
  explicit solution of the ground state.
\end{abstract}
\pacs{21.65.Qr, 12.38.Mh, 25.75.Nq}
\maketitle

\section{Introduction}

The quest for the phase diagram of strongly interacting matter out of
quarks and gluons (i.e.\ matter described by Quantum Chromodynamics --
QCD) is one of the most challenging problems in modern theoretical
and experimental physics.  There are many speculations on the QCD
phase diagram from theory such as the color-superconducting
phase~\cite{Alford:2007xm}, the quarkyonic
state~\cite{McLerran:2007qj} and the triple-point-like
structure~\cite{Andronic:2009gj}, the QCD critical
point~\cite{Asakawa:1989bq}, and so on (see
Refs.~\cite{MeyerOrtmanns:1996ea,Fukushima:2010bq} for comprehensive
reviews).  Available experimental information~\cite{Becattini:2005xt}
is, however, too limited to constrain uncertainties on those
speculative possibilities (see Ref.~\cite{Fukushima:2010is} for an
attempt and also Refs.~\cite{Cleymans:2005xv,Floerchinger:2012xd} for
physical interpretations).  Among others, the QCD critical point
search is vigorously ongoing in the present and future experimental
facilities as well as in the first-principle calculation of the
lattice-QCD simulation.

The QCD critical point would be, if discovered, a landmark for our
understanding on QCD matter.  In an infinite-volume system at
equilibrium, the fluctuations are expected to show critical behavior
and thus the criticality would serve as experimental
signatures~\cite{Stephanov:1998dy}.  There are many theoretical
proposals and experimental data taken from the beam-energy scan
program at Relativistic Heavy Ion Collider (RHIC) in the Brookhaven
National Laboratory.  It is an urgent question to make clear whether
the QCD critical point exists and, if any, where it is located.

Because of the notorious sign problem with finite quark chemical
potential $\muq$, the importance sampling breaks down in
finite-density simulations.  Although theoretical attempts are making
steady progresses, (temporal) finite-volume effects are not easily
treatable~\cite{Endrodi:2007gc}, and it is still difficult to extract
any reliable conclusion even on a qualitative level.  Then, under
these circumstances, there are three major passages toward the QCD
phase diagram studies (except for recent developments in the
functional method~\cite{Braun:2009gm}).

\paragraph*{1.}
One can discuss the critical phenomena \textit{assuming} the QCD
critical point.  This is a common strategy of theory in general.
Since the critical properties are universal, one can make
model-independent predictions.  The virtue of this approach is the
generality, but it does not give any clue about the concrete structure
of the QCD phase diagram.

\paragraph*{2.}
One can utilize the effective model description with a reasonable
choice of the model parameters~\cite{Ratti:2005jh}.  The location of
the critical point is sensitive to model details.  Not that all model
results are model dependent, but the nature of the phase transition at
high density strongly depends on a part of the model setup, as we will
elucidate later.

\paragraph*{3.}
One can make a conjecture on the phase structure based on generic
properties of QCD such as
symmetries~\cite{Pisarski:1983ms,Hatsuda:2006ps} and the degrees of
freedom in a particular limit~\cite{McLerran:2007qj}.  Because the
argument lacks for concrete dynamics unlike the model study, one
should check individually which scenario is favorable in reality.
Nevertheless, such a conjecture from physics deliberation provides us
with a useful guideline for model analysis.
\vspace{0.5em}

The aim of the present work is to establish a path from \textit{2} to
\textit{3} in the above classification.  This is a route rather
opposite to conventional approaches.  Instead of choosing a particular
model description, we shall extract the essential ingredients common in
most model studies and try to unveil the underlying physics mechanism
in a way free from model artifacts.  In particular, by looking at the
saturation curve, i.e.\ the energy per particle as a function of
density, we can clearly see the nature of the liquid-gas phase
transition, which also enables us to understand why the vector
interaction would disfavor the first-order phase transition.  It is a
straightforward extension to include inhomogeneity as the chiral
spiral for simplicity, and we can then find that the phase structure
still has rich contents, which is again understandable from the
saturation curve.

\section{First-order phase transition at zero temperature}

Let us start our analysis utilizing the same setup as
Ref.~\cite{Fukushima:2008is}.  We treat cold and dense quark matter in
a quasi-particle description.  This means that we assume a Fermi
liquid of quark matter, which should be valid for bulk thermodynamic
quantities as long as $T$ is small enough and the Landau damping is a
minor effect.  Strictly speaking, our strategy would work in a density
region between two onsets;  one for quark deconfinement and the other
for color superconductivity.  It is very hard to quantify
deconfinement and a phenomenological study of the equation of
state~\cite{Masuda:2012kf} implies that quark-hadron crossover may
start around the baryon density $\rho_{\rm B}\sim 2\rho_0$ with the
normal nuclear density $\rho_0\simeq 0.17\;\text{fm}^{-3}$.  The phase
structure involving color superconductivity is more ambiguous and
severely dependent on the models around
$\rho_{\rm B}\sim 5\rho_0$~\cite{Fukushima:2004zq}.  Therefore we
should restrict the validity of our treatment within a range
$2\rho_0 \lesssim \rho_{\rm B} \lesssim 5\rho_0$.  This is, however, a
rather conservative estimate and should be loosed at higher
temperature where quarks would be more liberated.

In this way the thermodynamic potential from quasi-particles,
$\Omega_{\text{matter}}$, is expressed as a function of the effective
mass $M$ in a form of
\begin{align}
 &\Omega_{\text{matter}}[M]/V = -\int_0^{\muq} d\mu' \rho(\mu') \notag\\
 &\qquad\quad -4\Nc\Nf\, T\int\frac{d^3 p}{(2\pi)^3} \ln\bigl(
  1+e^{-\omega_p/T} \bigr) \;,
\label{eq:Omega_matter}
\end{align}
where $\rho(\mu)$ is the quark number density defined by
$\rho(\mu)=2\Nc\Nf\int\frac{d^3 p}{(2\pi)^3}
[n_{\text{F}}(\omega_p-\mu) - n_{\text{F}}(\omega_p+\mu)]$ with the
Fermi-Dirac distribution function,
$n_{\text{F}}(\omega_p)=(e^{\omega_p/T}+1)^{-1}$, and the quasi-particle
energy, $\omega_p=\sqrt{p^2+M^2}$.  It is important to note that this
$\muq$-dependent matter part is common in any quark models such as the
(P)NJL and the (P)QM models~\cite{Ratti:2005jh}.  Then, the model
uncertainty is unavoidable in the vacuum part.

In a quasi-particle picture of quarks the vacuum part could be
expressed as $\Omega_0[M]/V = -2\Nc\Nf\int^\Lambda
\frac{d^3 p}{(2\pi)^3}\: \omega_p + U[M]$ with a potential term.  If
we postulate it as $U[M]=(M-m)^2/(4g_{\rm s})$, then
$\Omega_0[M]+\Omega_{\text{matter}}[M]$ exactly amounts to the
thermodynamic potential in the NJL model with the bare mass
$m$~\cite{Hatsuda:1994pi}.  To implement the $U(1)_{\rm A}$ anomaly in
the three-flavor case, we may add a term $-g_{\rm d}(M-m)^3$ in
$U[M]$.  From now on, we shall adopt a more general form of
$\Omega_0[M]$ inspired by the Ginzburg-Landau expansion, i.e.
\begin{equation}
 \Omega_0[M]/V = a(M_0^2-M^2)^2 -b M - c M^3\;.
\label{eq:GL}
\end{equation}
Although the thermodynamic potential in hand is extremely simple, this
setup sufficiently grasps the generic features of the phase transition
in cold and dense quark matter.  One may wonder if this polynomial
form would miss a logarithmic singularity as discussed in
Ref.~\cite{Skokov:2010sf}.  There are two reasons why this is not a
serious problem to our analysis:  First of all, such a logarithmic
singularity is related to the infrared singularity of massless fermion
loops.  As we will see later, we are more interested in the massive
case than the chiral limit and the effect of the logarithmic
singularity is only minor then.  Second, this logarithmic term has no
effect for the first-order phase transition at large $\muq$ and $T=0$
because the phase transition typically exists around $M \sim M_0$
(see Fig.~\ref{fig:chiral}), which is far from the singularity near
$M=0$.

To enter the regime at higher temperature, one should consider the
meson fluctuations that may give rise to $T$-dependent coefficients in
Eq.~\eqref{eq:GL}.  Therefore, strictly speaking, our analysis is
valid only in the region with $\muq\gg T$.  In what follows we
consider only the $c=0$ case, for we are interested in the mechanism
in favor of the first-order phase transition and $c\neq0$ would
trivially stabilize the first-order transition.

\begin{figure}
 \includegraphics[width=\columnwidth]{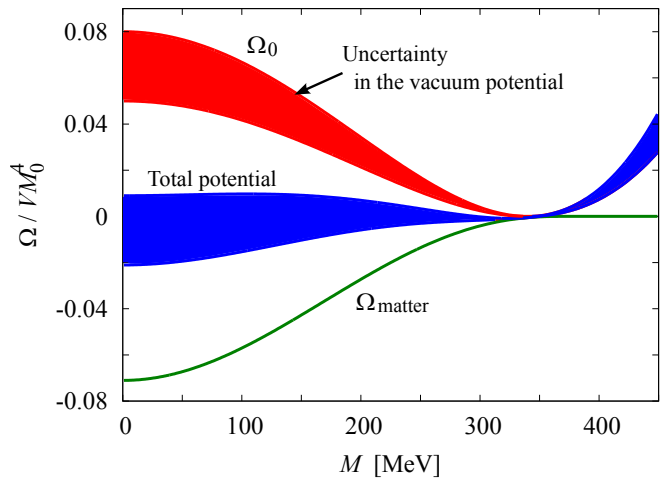}
 \caption{Potential shapes from Eqs.~\eqref{eq:Omega_matter} and
   \eqref{eq:GL} at $T=0$ with $\muq=370\;\text{MeV}$.
   $\Omega_{\text{matter}}$ is model independent, while the vacuum
   potential $\Omega_0$ leaves uncertainty.  The parameters in
   $\Omega_0$ are chosen as $M_0=340\;\text{MeV}$ and the curvature
   from $a=0.05$ ($\approx$ a value in the linear $\sigma$ model) to
   $a=0.08$ ($\approx$ a value in the NJL model), and $b=c=0$.}
 \label{fig:potential}
\end{figure}

Figure~\ref{fig:potential} shows the typical behavior of the potential.
As discussed in Ref.~\cite{Fukushima:2008is} the matter part
$\Omega_{\text{matter}}$ always has a minimum at $M=0$ because the
baryon density is the largest when quasi-particles are massless.  Let
us consider the condition for the first-order phase transition in the
case of $T=0$ in which Eq.~\eqref{eq:Omega_matter} simplifies as:
$\Omega_{\text{matter}}/V=-(\Nc\Nf/12\pi^2)\bigl(p_{\text{F}}\mu^3
-\frac{5}{2}M^2 p_{\text{F}}\mu+\frac{3}{4}M^4\ln[(\mu+p_{\text{F}})/
(\mu-p_{\text{F}})]\bigr)\,\theta(\mu-M)$ with
$p_{\text{F}}=\sqrt{\mu^2-M^2}$.  In Ref.~\cite{Fukushima:2008is} the
upper bound for the curvature $a$ was estimated under a reasonable but
limited situation, $\muq\simeq M_0$.  We can relax this numerically
only to find that a first-order phase transition can remain in the
chiral limit ($b=0$) unless we choose unphysical parameters so that a
phase transition takes place at $\muq\gg M_0$.  Then
$\Omega_{\text{matter}}$ stretches far beyond $M\sim M_0$ and the
phase transition is no longer of first order.

This simple analysis tells us that the first-order phase transition at
$T=0$ can occur since $\Omega_{\text{matter}}$ is proportional to
$\theta(\mu-M)$ and $\Omega$ does not have to contain a $M^6$ term,
while $\Omega$ is sometimes assumed to take a form of
$c_2 M^2+c_4 M^4+c_6 M^6$ at $T\neq0$.  Thus, the present formalism
based on the quasi-particle approximation is more appropriate for the
investigations of cold and dense quark matter.

Furthermore, we must add a term $\propto \rho^2$ in
$\Omega_{\text{matter}}$, which stems from the vector-channel
interaction~$(\bar{\psi}\gamma_\mu\psi)^2$ that is chiral
symmetric~\cite{Klimt:1990ws}, i.e.\
\begin{equation}
 \Omega_{\text{vec}}[M]/V = g_{\rm v}\rho^2 \;,
\label{eq:vector}
\end{equation}
which can be evaluated with $\rho$ numerically which is obtained as
$\rho=\frac{\Nc\Nf}{3\pi^2}\bigl(\mu^2-M^2\bigr)^{3/2}\theta(\mu-M)$
at $T=0$.  We should note that in the mean-field NJL model with the
vector interaction, usually, the vector interaction would shift the
chemical potential, which pushes the energy up by
$\sim 2g_{\rm v}\rho^2$, and the condensation energy is negative,
$-g_{\rm v}\rho^2$, leading to
$\sim 2g_{\rm v}\rho^2-g_{\rm v}\rho^2=g_{\rm v}\rho^2$ in total.
Here we simply postulate this in a form of Eq.~\eqref{eq:vector}.

\begin{figure}
 \includegraphics[width=\columnwidth]{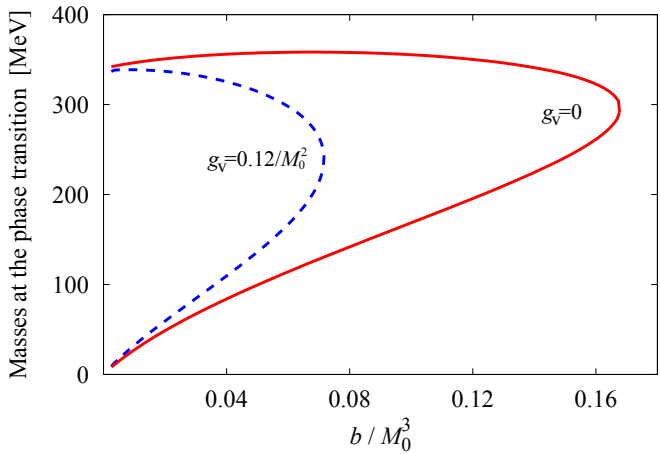}
 \caption{The location of the potential minima at the first-order
   transition point when $g_{\rm v}=0$ (solid curve) and
   $g_{\rm v}=0.12/M_0^2\simeq 10^{-6}\text{MeV}^{-2}$ (dashed
   curve) are chosen, respectively, with $a=0.05$ and
   $M_0=340\;\text{MeV}$.}
 \label{fig:first}
\end{figure}

For a deeper insight, Fig.~\ref{fig:first} is quite instructive.  This
figure shows the location of two degenerate minima in the potential
(i.e.\ the dynamical mass) when $\muq$ takes a value at the
first-order phase transition.  For example, in the chiral limit, the
dynamical quark mass jumps from $M\simeq M_0$ to $M=0$.  The jump is
naturally reduced at larger $b$ (larger quark mass) and eventually
only crossover remains beyond the bend of the curves in
Fig.~\ref{fig:first}.  One can notice that the curve substantially
shrinks with positive $g_{\rm v}$ which disfavors the first-order
phase transition.

It is interesting to see that the vector interaction has only a minor
impact for $b=0$.  This is because the minimum at $M=0$ is intact as
long as chiral symmetry is exact at $b=0$ and $\rho$ and thus the
vector interaction is still very small at $M=M_0$.  This observation
is, however, not completely free from the model choice.  If the phase
transition is located at $\muq > M_0$ with some other choice of
parameters, the potential minimum around $M=M_0$ is also influenced
substantially by the density effect and thus the first-order phase
transition could be diminished by the vector interaction.  This part
of uncertainty is not relevant, for we are interested in the physical
world with finite quark mass after all.

Guided by Fig.~\ref{fig:first} we shall specifically look at the
following three cases:  (1) $b=g_{\rm v}=0$ (first-order),
(2) $b=0.08 M_0^3$ and $g_{\rm v}=0$ (weak first-order), and (3)
$b=0.08 M_0^3$ and $g_{\rm v}=0.12/M_0^2$ (crossover).

\begin{figure}
 \includegraphics[width=\columnwidth]{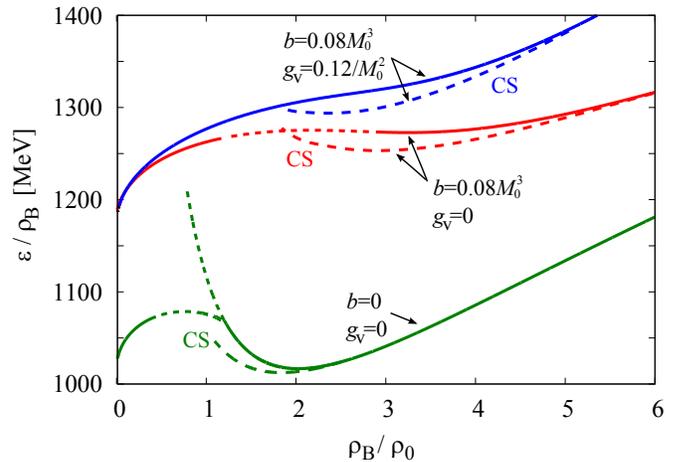}
 \caption{Energy per particle as a function of the density.  The solid
 curves represent the homogeneous results for (1) $b=g_{\rm v}=0$, (2)
 $b=0.08 M_0^3$ and $g_{\rm v}=0$, (3) $b=0.08 M_0^3$ and
 $g_{\rm v}=0.12/M_0^2$ from the bottom to the top.  The dashed curves
 with the label ``CS'' represent the chiral-spiral results for
 respective parameters.  The horizontal axis is given in the unit of
 the normal nuclear density $\rho_0=0.17\;\text{fm}^{-3}$.}
 \label{fig:energy}
\end{figure}

For later convenience we shall plot the energy per particle
$\varepsilon/\rho_{\rm B}$ at $T=0$ in Fig.~\ref{fig:energy}, where
$\varepsilon=\Omega/V+\mu_{\rm B}\rho_{\rm B} - \Omega_0/V$ is the
internal energy density measured from the hadronic vacuum with
$M\sim M_0$ (before a finite density appears), and
$\rho_{\rm B}=\rho/\Nc$ is the baryon number density.  If the curve
has a minimum as a function of $\rho_{\rm B}$, i.e.\
$d(\varepsilon/\rho_{\rm B})/d\rho_{\rm B}
=\mu_{\rm B}/\rho_{\rm B}-\varepsilon/\rho_{\rm B}^2=0$, the pressure
difference becomes zero, which indicates a first-order phase
transition of the general liquid-gas type (see
Ref.~\cite{Tatsumi:2011tt} for a review and also
Ref.~\cite{Chomaz:2004nw} for experimental studies).  Therefore,
whenever $\varepsilon/\rho_{\rm B}$ has a minimum as a function of
$\rho_{\rm B}$, the $T=0$ system must have a first-order phase
transition in the same way as the (symmetric) nuclear matter phase
transition at $\mu_{\rm B}=M_N-B$ with $M_N\simeq 939\;\text{MeV}$
being the nucleon mass and $B\simeq 16\;\text{MeV}$ the nuclear
binding energy.  At the second-order transition, the energy
curve should be flat at the point of inflection.  This kind of
analysis on quark matter is well known in the context of quark
droplets~\cite{Buballa:1996tm} but less applied in the phase diagram
research.  What is necessary for the existence of the critical point
(first-order phase transition) is a convex structure of the curve
(saturation property), which is a general statement that does not rely
on any model nor Ansatz.

\begin{figure}
 \includegraphics[width=0.7\columnwidth]{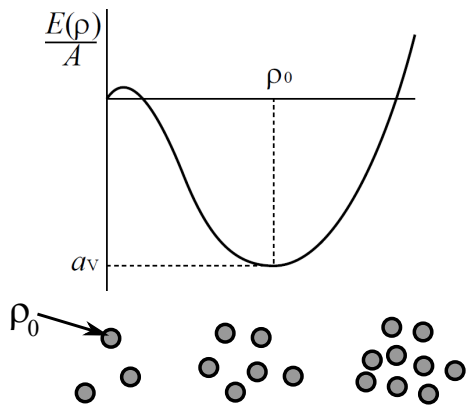}
 \caption{Schematic figure of the saturation curve of nuclear matter
   with a minimum at $\rho_0=0.17\;\text{fm}^{-3}$ and the binding
   energy given by the volume term $a_V$ in the Bethe-Weizs\"{a}cker
   mass formula.  An intermediate density $\rho<\rho_0$ can be
   realized as a spatial average over bubbles with the core with
   $\rho\sim\rho_0$ in the empty vacuum.  Though the surface energy
   effect is not considered in the above schematic figure where a
   simple nucleon-gas picture is depicted, the actual bubble shapes in
   a nuclear liquid depend on the surface term $a_S$, etc.}
 \label{fig:saturation}
\end{figure}

Because this point of the liquid-gas transition is so important, let
us recall here how an intermediate density between $\rho=0$ and the
saturation density $\rho=\rho_0$ can be realized in this case.  If the
energy per particle has a minimum as schematically shown in the upper
panel of Fig.~\ref{fig:saturation} it would be energetically
preferable to form bubbles with the core with $\rho\sim\rho_0$
rather than a homogeneous distribution of dilute $\rho$.  If we
consider the surface energy, the density gradient (Weizs\"{a}cker)
term, and the charge neutrality, bubbles should take optimal shapes
such as the nuclear pasta (spaghetti, lasagna,
etc)~\cite{Watanabe:2000rj}.  Such a state of matter is nothing but a
mixed phase associated with the first-order phase transition, and
importantly, this argument already implies the existence of an
inhomogeneous ground state near the liquid-gas transition.  In other
words, if a mixed phase is characterized by a typical wave number $q$,
how can we strictly distinguish such a phase from an inhomogeneous
ground state?  One may think that in the case of quark matter the
inhomogeneity is turned on not in the density only but in the mass
$M$ unlike nuclear matter.  We would stress, however, that $M$ also
controls the density and the physics is just the same if seen in terms
of the saturation curve as in Fig.~\ref{fig:energy}.

It is obvious from Fig.~\ref{fig:energy} that the vector interaction
as in Eq.~\eqref{eq:vector} disfavors the first-order phase
transition.  The minimum in $\varepsilon/\rho_{\rm B}$ is pushed up by
the quadratic term $\propto \rho_{\rm B}^2$ and eventually the
first-order phase transition disappears when the minimum is lost, as
demonstrated by three solid curves in Fig.~\ref{fig:energy}.  In the
chiral limit $b=0$ the branch of $M=0$ is separate, so that the
first-order phase transition survives regardless of the vector
interaction, which may change with different parameters as we already
pointed out.  With finite $b$, however, two branches with small and
large $M$ are smoothly connected and the minimum diminishes for large
$b$ and $g_{\rm v}$ in accord to Fig.~\ref{fig:first}.

\section{Chiral spirals}

One may find the usefulness of the saturation curve for analyses with
a wider range of model space.  From now on we shall consider the
possibility to form inhomogeneous chiral condensates.  We here utilize
the simplest Ansatz to introduce it, namely, the one-dimensional
chiral spiral;
$\langle\bar{\psi}\psi\rangle=\chi\cos(2qz)$ and
$\langle\bar{\psi}\gamma_5\tau_3\psi\rangle=\chi\sin(2qz)$
(see Ref.~\cite{Schon:2000qy} for reviews).  This ground state of the
chiral spiral can be equivalently described by a chiral rotation
$\psi=e^{i\gamma_5 \tau_3 q z}\psi'$ with a homogeneous condensate
$\chi=\langle\bar{\psi}'\psi'\rangle$ in the chiral limit.  Then, the
quasi-particle dispersion relation in the $\psi'$-basis is expressed
as~\cite{Schon:2000qy,Migdal:1978az}
\begin{equation}
 \tilde{\omega}_p = \sqrt{p_\perp^2+(\sqrt{p_z^2+M^2}\pm q)^2} \;,
\label{eq:disp_q}
\end{equation}
where $\pm$ in front of $q$ corresponds to the flavor and the
chirality that also depends on the sign of $p_z$.

This type of inhomogeneity pattern has been considered repeatedly in
various contexts such as the pion condensation in nuclear
matter~\cite{Migdal:1978az}, large-$\Nc$ QCD~\cite{Deryagin:1992rw},
the Overhauser instability~\cite{Nakano:2004cd}, the quarkyonic spiral
with confining force~\cite{Kojo:2009ha}, and so on.  The dispersion
relation~\eqref{eq:disp_q} should be plugged into
$\Omega_{\text{matter}}/V$ in Eq.~\eqref{eq:Omega_matter}.  Unlike the
normal dispersion relation, we see that a large part of the mass
effect can be absorbed by $q\sim M$, with which $\rho$ is no longer
suppressed even at large $M$.  This is the reason why a first-order
phase transition can occur from the homogeneous hadronic phase to the
chiral spiral where $M$ is substantially large.  Also, we should point
out that the Ginzburg-Landau analysis in Ref.~\cite{Carignano:2010ac}
to conclude that the chiral spiral is less favored might be
inadequate;  the largest energy gain in $\Omega_{\text{matter}}/V$
comes from the region with large $M$ where the Ginzburg-Landau
expansion should not work.

The physical mechanism to lower the total energy is the Overhauser
effect as argued in Ref.~\cite{Nakano:2004cd}.  In the ordinary
Overhauser instability the momenta of the spin-up component are shifted
up by $p_{\text{F}}$ and those of the spin-down component are shifted
down by $p_{\text{F}}$, so that a gap opens where two energy
dispersion relations cross.  In (1+1)-dimensional NJL model the
situation is completely analogous~\cite{Schon:2000qy};  a choice of
$q=2\muq$ eliminates the $\muq$ dependence and the energy gain
originates from the fact that $\rho$ is completely insensitive to $M$
and thus $\rho$ is never suppressed by $M$ in contrast to the
homogeneous solution.  In (3+1)-dimensional case, on the other hand,
not only $p_z$ but also $p_\perp$ share the Fermi momentum, and so the
optimal $q$ is not $2\muq$ but rather $q\sim M$ which will be
confirmed by numerical calculations later.

Thus, $\Omega_{\rm matter}$ always tends to favor the chiral spiral
with $q\sim M$, while it is $\Omega_0$ that would hinder the growth of
$q$.  In the leading order the vacuum part has an expansion in terms
of $q$ as
\begin{equation}
 \Omega_0[M,q]/V = \Omega_0[M,q=0]/V + (\alpha M^2 + \beta b) q^2 \;,
\end{equation}
where the first term with $\alpha>0$ is a ``kinetic'' term against
spatial modulation.  This term should be vanishing at either $M=0$ or
$q=0$, so the expansion should start with $M^2 q^2$.  One can estimate
$\alpha$ using a chiral model, but one should be careful not to pick
an unphysical term $\sim\Lambda^2 q^2$ up from gauge-variant
regularization.  The latter term $\propto \beta$ comes from a phase of
the current mass term associated with the basis change from $\psi$ to
$\psi'$.  Quantitative details may depend on $\alpha$ and $\beta$, but
qualitative features as we discuss below do not rely on any specific
choice of them.

\begin{figure}
 \includegraphics[width=\columnwidth]{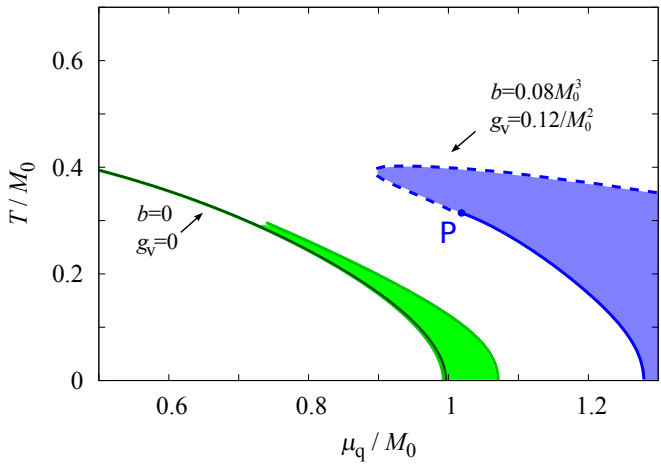}
 \caption{Typical phase diagrams with the chiral spiral.  The solid
   curve in the lower-$\muq$ side represents the homogeneous chiral
   phase transition of first order with $b=g_{\rm v}=0$ with which the
   chiral spiral region surrounded by the first-order phase boundaries
   is attached.  For $b=0.08M_0^3$ and $g_{\rm v}=0.12/M_0^2$ the
   homogeneous first-order transition and thus the QCD critical point
   no longer appear, but the inhomogeneous region is enlarged as shown
   in the higher-$\muq$ side with a first-order boundary (solid curve)
   terminating at $\textsf{P}$ followed by a second-order boundary
   (dashed curve).}
 \label{fig:diagram}
\end{figure}

\begin{figure}
 \includegraphics[width=\columnwidth]{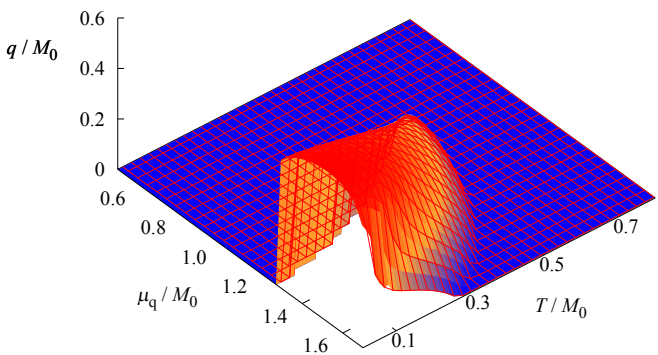}
 \caption{Behavior of the wave-number $q$ as a function of $\muq$ and
   $T$ in the unit of $M_0$ in the case with $b=0.08M_0^3$ and
   $g_{\rm v}=0.12/M_0^2$.  This clearly shows the structure of the
   chiral-spiral island, surrounded by a steep ``cliff'' at smaller
   $\muq$ and a gentle ``beach'' at larger $\muq$.}
 \label{fig:wavenum}
\end{figure}

Figure~\ref{fig:diagram} shows typical behavior of the phase
boundaries on the $\muq$-$T$ plane with zero and non-zero $b$ and
$g_{\rm v}$.  For demonstration we chose $\alpha=0.25$ and
$\beta=0.25/M_0$.  Then in the lower-$\muq$ side of
Fig.~\ref{fig:diagram} we see that there is an island structure of the
chiral spiral surrounded by the first-order boundaries.  The solid
curve extending to smaller $\muq$ represents a first-order phase
transition associated with the homogeneous condensate only.  It should
be mentioned that the first-order phase transition at $b=0$ in the
high-$T$ and small-$\muq$ region, which is not of our present
interest, might have been artificially strengthened due to the lack of
the logarithmic singularity in Eq.~\eqref{eq:GL}.  The first-order
boundary of inhomogeneity at smaller $\muq$ stays very close to this
curve.  This is because the effective potential becomes very shallow
near the first-order phase transition in the homogeneous case as
clearly recognized in the total potential presented in
Fig.~\ref{fig:potential}.  The secondary first-order boundary at
larger $\muq$ is much weaker because $M$ and thus $q$ are small there.
(Note that, in the chiral limit, $q$ may not decrease but only
increase in a narrow region of $\muq$ as shown in
Ref.~\cite{Nakano:2004cd}.  This tendency near the first-order phase
transition is partially seen also in the massive case in
Fig.~\ref{fig:wavenum}.)  The corresponding saturation curve of
$\varepsilon/\rho_{\rm B}$ is shown by a long-dashed curve with the
label ``CS'' in Fig.~\ref{fig:energy}, from which a minimum at lower
energy is apparent.  We note that the inhomogeneity island in the
vicinity of the first-order phase transition is consistent with our
intuitive discussions of the mix phase formation below
Fig.~\ref{fig:saturation}.

With the vector interaction included, the so-called QCD critical point
is easily washed
out~\cite{Fukushima:2008is,Carignano:2010ac,Kitazawa:2002bc}.
Interestingly, however, as shown in the higher-$\muq$ side of
Fig.~\ref{fig:diagram} and especially at \textsf{P} in this figure,
there is a chance that the critical point (strictly speaking,
tri-critical point) is revived driven by the inhomogeneous condensate.
The question is then how robust this observation is.  In fact it has
been reported that the soliton solution~\cite{Nickel:2009wj} is more
stable than the chiral spiral and also it exhibits a second-order
phase transition rather than a first-order
one~\cite{Carignano:2010ac}.

Let us then consider when the second-order phase transition is
possible in view of the saturation curve in Fig.~\ref{fig:energy}.  To
have a second-order phase transition from the hadronic phase (with
homogeneous $M\sim M_0$) to a general inhomogeneous state, there must
be an energy curve that is tangent to the hadronic branch (solid
curves from $\rho_{\rm B}=0$) and going below it.  The curves do not
have to be flat because there is a small energy difference before and
after a finite density appears, which is further enhanced by
$1/\rho_{\rm B}^2$ in the slope of the saturation curve.  To avoid a
first-order transition, moreover, the energy curve should be
monotonically increasing with increasing $\rho_{\rm B}$.

Such a situation is not allowed, for example, in the far bottom curves
(at $b=g_{\rm v}=0$) in Fig.~\ref{fig:energy}.  In this case with the
saturation energy lower than that at $\rho_{\rm B}=0$, we can conclude
that only a first-order phase transition is possible however
complicated and optimized modulations we introduce.  The situation is
different with finite $b$ and/or $g_{\rm v}$.  It is clear on a
qualitative level that a larger $g_{\rm v}$ would ease better
inhomogeneous states to develop, for the dashed chiral-spiral curve
could be then easily extended down to $\rho_{\rm B}=0$ monotonically.
This means that the phase transition between the homogeneous and
inhomogeneous states could be of second order.  Therefore,
unfortunately, the existence of the critical point \textsf{P} is again
not a robust conclusion especially with the vector interaction.

From a plain physical interpretation, it would be the most natural to
have continuous phase transitions that border the inhomogeneous
island.  Such an intuition is based on the picture of the liquid-gas
phase transition.  In fact, if the boundary is a first-order phase
transition, there will appear a density regime that can be described
only as a mixed state.  It is the role of the chiral condensate in
quark matter that makes a difference from the situation in nuclear
mater.  The density modulation inherent in a mixed state can be
mimicked by the modulation in the chiral condensate, which would lead
to an inhomogeneous ground state of quark matter with lower energy.
This is exactly what happens with the soliton solution in
Refs.~\cite{Nickel:2009wj,Carignano:2010ac}.  Indeed, at the onset of
solitonic inhomogeneity, localized domain-walls start appearing,
which approaches sinusoidal patterns at larger $\muq$.  The density
profile has peaks arising from the kinks and this situation is
reminiscent of a mixed state picture as schematically depicted in the
bottom of Fig.~\ref{fig:saturation}.  It would interesting to figure
out the saturation curve corresponding to the solitonic solution.
This is beyond our current scope, but it presumably goes below the
chiral-spiral curves and is smoothly merged with the hadronic branch
at smaller $\rho_{\rm B}$.

\begin{figure}
 \includegraphics[width=\columnwidth]{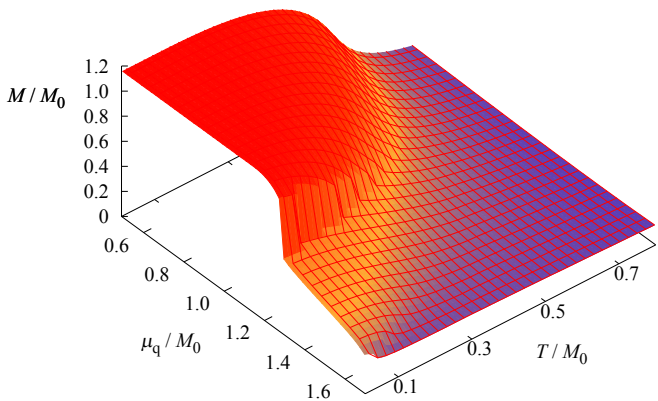}
 \caption{Behavior of the constituent mass $M$ as a function of $\muq$
   and $T$ in the unit of $M_0$ in the case with $b=0.08M_0^3$ and
   $g_{\rm v}=0.12/M_0^2$.  At the phase boundary into the chiral
   spiral, $M$ drops but still remains finite, and goes smaller
   continuously with increasing $\muq$ and/or $T$.}
 \label{fig:chiral}
\end{figure}

Finally let us take a closer look at the solution with $b=0.08M_0^3$
and $g_{\rm v}=0.12/M_0^2$.  Figures~\ref{fig:chiral} and
\ref{fig:wavenum} show the behavior of the constituent mass $M$ and
the wave-number $q$, respectively, as functions of $\muq$ and $T$.
At a glance one may notice that  $q\sim M$ holds in the chiral-spiral
region as we discussed.  The structure of the chiral-spiral island is
quite characteristic.  In view of Fig.~\ref{fig:wavenum}, one might
say that the island is surrounded by a steep ``cliff'' at smaller
$\muq$ and a gentle ``beach'' at larger $\muq$~\footnote{This is a
  concrete manifestation of the ``Happy Island'' conjectured by
  Larry~McLerran~\cite{McLerran:2011zj}.}.

Such a structure of the island should be quite robust because the
energy gain is mainly attributed to $q\sim M$.  Hence, the cliff
stands with a large energy gain at smaller $\muq$ where $M$ is still
large, and the inhomogeneous state gradually becomes indistinguishable
from the homogeneous state as $M$ gets smaller at larger $\muq$.

\section{Summary}

We have developed a picture of the first-order phase transition of
quark matter based on the saturation curve and the liquid-gas phase
transition.  From this picture we discuss the relation between the
order of the phase transition and the behavior of the saturation
curve.  We demonstrated this using a simple Ansatz of the chiral
spiral, but the argument itself is not limited to such a special
choice.  As a matter of fact, because the chiral spiral can be mapped
to the conventional pion condensation~\cite{Migdal:1978az} that is
killed by the spin-isospin interaction, it may be likely that the
chiral spiral should be suppressed by the axial-vector interactions
$\sim (\bar{\psi}\gamma_5\gamma_\mu\psi)^2$ or
$\sim (\bar{\psi}\gamma_5\gamma_\mu\vec{\tau}\psi)^2$, and eventually
superseded by others such as the soliton-like modulation and more
generally multiple-wave superpositions.  Even in this case the
saturation curve would provide us with valuable information on the
nature of the phase transition.

We can think of several directions as future extensions.  It may be
interesting to seek for some connections between our saturation
considerations and the Ginzburg-Landau analyses as in
Ref.~\cite{Abuki:2011pf}.  Also, the interplay between the chiral
spiral and the external magnetic field would deserve further
investigations~\cite{Basar:2010zd}.  We are actually working in this
direction to clarify the phase structure with three axes, $\muq$, $T$,
and $B$, including the spatially inhomogeneous state~\cite{future}.


\acknowledgments
The author thanks Tetsuo~Hatsuda, Yoshimasa~Hidaka, Teiji~Kunihiro,
Larry~McLerran, Toshiki~Tatsumi, and Wolfram~Weise for critical
comments and useful discussions.


\end{document}